\documentclass[12pt,preprint]{aastex}

\usepackage{color}

\shorttitle{Rotation and nitrogen enrichment in massive stars}
\shortauthors{Hunter et al.}

\begin{document}

\title{The VLT-FLAMES survey of massive stars: rotation and nitrogen 
enrichment as the key to
understanding massive star evolution}

\author{I. Hunter\altaffilmark{1,2}, I. Brott\altaffilmark{3}, 
D.J. Lennon\altaffilmark{4}, 
N. Langer\altaffilmark{3}, P.L. Dufton\altaffilmark{1}, 
C. Trundle\altaffilmark{1}, S.J. Smartt\altaffilmark{1}, 
A. de Koter\altaffilmark{5},
C.J. Evans\altaffilmark{6} and R.S.I. Ryans\altaffilmark{1}}%

\altaffiltext{1}{Astrophysical Research Centre, School of Mathematics and
Physics, The Queen's University of Belfast, Belfast, BT7 1NN, 
Northern Ireland, UK}
\altaffiltext{2}{The Isaac Newton Group of Telescopes, Apartado de Correos 321, 38700 
Santa Cruz de La Palma, Canary Islands, Spain}
\altaffiltext{3}{Astronomical Institute, Utrecht University, Princetonplein 5,
NL-3584CC, Utrecht, Netherlands}
\altaffiltext{4}{Instituto de Astrof\'{i}sica de Canarias, Calle V\'{i}a L\'{a}ctea,
E-38200 La Laguna, Spain}
\altaffiltext{6}{Astronomical Institute Anton Pannekoek, University of Amsterdam, Kruislaan 403, 1098
SJ Amsterdam, Netherlands}
\altaffiltext{5}{UK Astronomy Technology Centre, Royal Observatory, Blackford Hill, 
Edinburgh, EH9 3HJ}

\begin{abstract}
Rotation has become an important element in evolutionary models
of massive stars, specifically via the prediction of rotational mixing. 
Here, we study a sample of stars, including rapid rotators, to constrain
such models and use nitrogen enrichments as a probe of the mixing
process. Chemical compositions (C, N, O, Mg and Si) have been estimated for 135
early B-type stars in the Large Magellanic Cloud with projected 
rotational velocities up to $\sim$300\,km\,s$^{-1}$ using a non-LTE {\sc
tlusty} model atmosphere grid. 
Evolutionary models, including rotational mixing, have been generated 
attempting to reproduce these observations by adjusting the overshooting and
rotational mixing parameters and produce reasonable agreement with
60\% of our core hydrogen burning sample. 
We find (excluding known binaries) a significant population
of highly nitrogen enriched intrinsic slow rotators 
($v \sin i \lesssim$50\,km\,s$^{-1}$) incompatible with our models ($\sim$20\% of the sample). 
Furthermore, while we find fast rotators with enrichments
in agreement with the models, the observation of evolved ($\log g < 3.7$dex)
fast rotators that are relatively unenriched (a further $\sim$20\% of the sample)
challenges the
concept of rotational mixing.  We also find that 70\% of our blue supergiant sample
cannot have evolved directly from the hydrogen burning main-sequence.
We are left with a picture where invoking binarity and perhaps fossil magnetic fields
are required to understand the surface properties of a population of massive main sequence stars.

\end{abstract}


\section{Introduction} \label{s_intro}

Stellar rotation in massive stars has been adopted in recent theoretical
models to, for example, predict the correct blue to red supergiant ratio \citep{mae01},
the progenitors of gamma-ray bursts through homogeneous evolution \citep{yoo05} and 
Wolf-Rayet populations as a function of metallicity (\citealt{mey05}; \citealt{vin05}). 
However, the consequent
surface enrichment of helium and nitrogen through rotational mixing is poorly constrained
by observations \citep{daf01}, with the mixing typically calibrated from 
either studies of main-sequence
stars with low projected rotational velocities \citep{kor02}
or evolved supergiant stars \citep{ven99}. The former are biased towards slow rotators while the latter
have evolved beyond the core hydrogen burning phase. The analysis 
of an unbiased sample of fast rotating core-hydrogen
burning stars in order to properly calibrate the predicted rotational mixing
efficiency was a strong driver for
a large survey of O- and
early B-type stars in our 
Galaxy and the Magellanic Clouds
\citep{eva05}. \citet{hun07a} and \citet{tru07}, hereafter Paper I and II 
respectively,
have presented chemical
abundances for the narrow lined objects (predominately slow rotators) 
in the LMC sample. Here we extend the study
with the chemical analysis of the faster rotating stars.

\section{Observations and model atmosphere analysis}\label{s_obs}

High-resolution ($R\sim20\,000$) 
spectra from the Fibre Large Array Multi-Element Spectrograph (FLAMES) at the European
Southern Observatory Very Large Telescope 
were obtained for some 750 O- and early B-type stars located towards clusters
in our Galaxy and the Large and Small Magellanic Clouds (LMC and SMC respectively).
A discussion of target selection, observational details and initial data reduction
can be found in \citet{eva05}. In this letter we discuss the LMC early B-type 
stars, using nitrogen as a probe of chemical enrichment.

A grid of non-LTE {\sc TLUSTY} model atmospheres  
\citep{hub95} has been used to calculate the
atmospheric parameters and chemical compositions (C, N, O, Mg and Si) of our 
targets as described
in Papers I, II and \citet{hun07b}, hereafter Paper III.
We have fitted the lines with
rotationally broadened profiles to estimate the equivalent widths (EWs), since at
significant rotational velocities ($>$50\,km\,s$^{-1}$) the line shape is rotationally dominated.
For those objects in which no nitrogen features could be identified, an upper limit
to the nitrogen abundance was estimated by placing an upper limit on the equivalent width
of the strongest N\,{\sc ii} line in our spectral range, which is located at 3995\AA\ (Paper I).

The mean abundances of C, O, Mg and Si are in excellent agreement (within 0.05\,dex) 
with the LMC baseline abundances given in Paper I. However the mean nitrogen abundance is
0.36\,dex higher than its baseline abundance indicating that nitrogen enrichment
has occurred in many of the stars. 
Additionally the scatter in the nitrogen abundances is over a factor of two larger
than that of the other elements, indicating that different levels of enrichment explain the
mean nitrogen abundance, rather than 
systematic errors. Since nitrogen is an important element in the CNO-cycle,
the surface nitrogen enrichments can be used as a measure of the mixing efficiency. 
Note that although a corresponding carbon depletion
of up to $\sim$0.2\,dex
would be expected, within the uncertainties in determining carbon abundances
(see Paper I), such
an effect would be difficult to observe.
In Fig.~\ref{f_nvsini}
the nitrogen abundances are plotted as a function of the projected rotational velocity with
the sample being split into two groups; core hydrogen burning objects with 
surface gravities $\geq$3.2\,dex and supergiants
 having surface gravities $<$3.2\,dex (see Paper III). 

\section{Stellar evolution models} \label{s_models}

Comparison with published stellar evolution models such as the Geneva models \citep{mae01} 
is complicated as no rotating models at the LMC metallicity are currently available.
Additionally, the initial chemical composition for evolutionary models is often scaled from 
solar composition, and this is not appropriate for all elements, in particular nitrogen.

New evolutionary models have been calculated (Brott et al. in prep.) 
using the stellar evolution code of \citet{yoo06}, 
which includes rotation 
\citep{heg00b} and angular momentum transport by magnetic torques \citep{spr02}. 
Only two differences apply here:
we updated the mass loss prescription and now use the recipe of \citet{vin01}, and we disregard 
the magnetically induced chemical diffusion term of \citep{spr02},
which --- in contrast to the magnetic angular momentum transport --- is not
observationally supported, and appears controvercial at present \citep{spr06}. 
 

As initial composition we adopted the LMC C, N, O, Mg and Si abundances given in Paper I and
all the other metal abundances decreased by 0.4\,dex from the solar composition of \citet{asp05}.
Based on the recent discussion of the primordial helium abundance \citep{pei07},
we have updated the initial helium mass fraction for our LMC models to Y= 25.6\%, which together
with the metallicity of the adopted chemical composition (Z= 0.5\%), results in a hydrogen
mass fraction of X= 73.9\%.

In Paper III we determined
that the end of core hydrogen burning occurs at a logarithmic surface gravity of 
$\sim$3.2\,dex. We find that a convective core overshooting of 0.335 
pressure scale heights is required to reproduce this result.  While this value
is larger than what is typically assumed, with consequences 
for the post-main sequence
evolution which still remain to be investigated, it provides the only way to 
understand the sharp drop in the rotational velocity 
distribution of our sample at a surface gravity of 3.2\,dex. 
Evolutionary models neglecting overshooting
indicate that the surface nitrogen
enrichment is smaller by less than  0.15\,dex at core hydrogen
exhaustion for a rapid rotator. Hence, our principal conclusions
are not significantly affected by the adopted amount of overshooting.

The efficiency of rotationally induced mixing in our models 
is controlled by the parameter $f_{\rm c}$, 
which is the ratio 
of the turbulent viscosity to the diffusion coefficient that describes the transport 
of angular momentum
by rotationally induced hydrodynamic 
instabilities \citep{heg00b}. The mean projected rotational velocity 
of the stars shown in Fig.~\ref{f_nvsini} is 110\,km\,s$^{-1}$, 
with the mean surface nitrogen abundance being 7.2\,dex.
Although this mean 
projected rotational velocity is lower than that observed for cluster stars
(see, for example, \citealt{hua06}) there is no bias towards stars with low
rotational velocities. Our stars are a relatively unbiased sample of the true populations,
as explained in \citet{eva06}. This population is dominated by field stars, which are
known to rotate slower than cluster stars, and our velocity distribution is
comparable to the Galactic field stars of \citet{abt02}. Additionally Be-type stars
are excluded, hence our mean velocity is representative only for normal
B-type stars and there is no velocity bias for these B-type stars.

The parameter $f_{\rm c}$ has been calibrated to reproduce the mean 
surface nitrogen abundance of the core-hydrogen burning stars
at core hydrogen exhaustion for a 13\,M$_\sun$ model (the mean mass of our non-supergiant stars)
initially rotating at 140\,km\,s$^{-1}$ (110$\times$4/$\pi$ to account for random angles of inclination).
We find $f_{\rm c}$=2.28$\times$10$^{-2}$, compared to 3.33$\times$10$^{-2}$ adopted previously 
(\citealt{heg00b}). Using these parameters, we computed a grid of models for masses
representative of our sample and a range of initial rotational velocities.

\section{Discussion and conclusions}

\begin{figure*}[ht]
\plotone{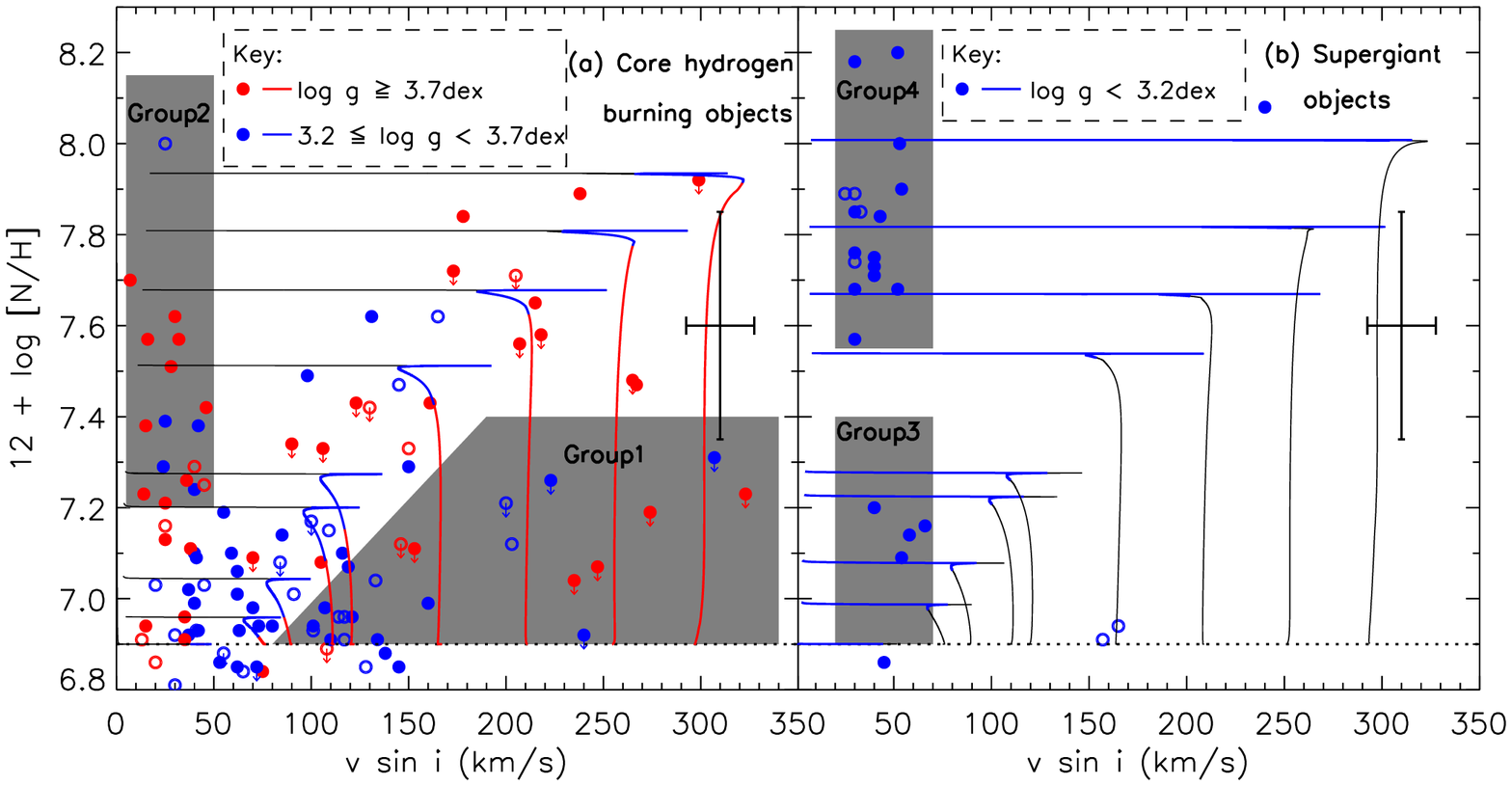}
\caption{Nitrogen abundance (12 + log [N/H]) against the projected 
rotational velocity ($v\sin i$) for core hydrogen burning (a) and supergiant (b) objects. 
Open symbols: radial velocity variables; downward arrows: abundance upper limits;
dotted line: LMC baseline nitrogen
abundance.  The mean uncertainty in the nitrogen abundance is 0.25\,dex while that in $v\sin i$ is
10\% and these are illustrated. These errors are largely independant
of rotational velocity since the systematic uncertainties are comparable to the 
measurement errors. 
The bulk of the core hydrogen burning objects occupy a region at low 
$v\sin i$ and show little or modest nitrogen enrichment.
The tracks are computed 
for an initial mass of 13\,M$_\sun$ (a) and 19\,M$_\sun$ (b),
corresponding to the average mass of our non-supergiant and supergiant stars, respectively,
and their rotational velocity has been multiplied by $\pi$/4.  Although the plot
contains stars with a range of masses, comparison of the tracks shown in (a) and (b)
show that any mass effect is negligible compared to the abundance uncertainties.
In (a) the surface gravity has been used as indicator of the evolutionary status and the objects
(see legend) and tracks have been split into red and blue to 
indicate younger and older stars respectively.  However, this is
illustrative only, since the evolutionary status is obviously continuous and not discrete.
Gray shading in panel~(a) highlights two groups of stars which remain unexplained
by the stellar evolution tracks. In panel~(b), gray shading highlights the apparent
division of the supergiants into two distinct groups. The
surface gravity estimates of many
of the objects in Group~3, and the two apparantly rapidly rotating unenriched
supergiants,  are consistent
with being in the core hydrogen phase within their uncertainties. }

\label{f_nvsini}
\end{figure*}

In Fig.~\ref{f_nvsini} the nitrogen abundances are plotted against
the projected rotational velocities for our sample stars, and compared with the evolutionary models.
Given the large number of objects in our sample, it is reasonable to assume a 
random distribution of inclination angles,
and hence the rotational velocity of the tracks has been scaled by $\pi$/4.
Although we have attempted to contrain the evolutionary
tracks to the observations, we are able to reproduce agreement for only 60\% of the data
(excluding known radial velocity variables).


However, two groups of core hydrogen burning stars in Fig.~\ref{f_nvsini}(a)
stand out as being in conflict with the evolutionary models. 
Group~1 contains fast rotators which have little chemical mixing. 
In particular, this group includes fast rotating single stars with  
surface gravities indicating that they are near the end of core hydrogen burning (excluding radial
velocity variables this is $\sim$20\% of the
core hydrogen burning sample.) 
The gravities used to discriminate these stars have not been
corrected for the fact that we may observed them almost equator-on, i.e.
their true polar gravities could be larger, meaning that they are less evolved than our
derived gravities imply. However, applying the surface gravity corrections described in \citet{hua06b}
would increase the gravities by up to $\sim$0.3\,dex for our faster rotators. 
As the zero age main-sequence
gravity of our 13\,M$_{\sun}$ models is about 4.3\,dex, and since most
of the stellar expansion on the main sequence occurs towards the end of core
hydrogen burning, this gravity uncertainty can not reconcile the situation.


The second discrepant group of stars in Fig.~1a are the 15 slow rotators 
($v \sin i < $50\,km\,s$^{-1}$) that show significant nitrogen enrichment (Group~2). This group
also forms $\sim$20\% of the non-binary core hydrogen burning sample.
Although it could be argued that they may be fast rotators observed
pole-on, statistically this is unlikely.
In order to reproduce their mean nitrogen abundance requires
a rotational velocity of $\sim$200\,km\,s$^{-1}$. 
Using simple geometry to calculate the solid angle 
restriction that we must impose, i.e. $4\pi(1-cos\theta)$,
and assuming that we need 15 stars
to have this velocity, we expect less than one star to appear with
such a low rotational velocity due to 
random inclinations. 
The simple corollary of this is that if we have 15 stars with
velocities $\sim$200\,km\,s$^{-1}$ populating 
Group~2 through random angles of inclination, over 300 
similary enriched stars with $v\sin i$ values between 150 and 200\,km\,s$^{-1}$ would be expected.
Such a population clearly does not exist.
 
Evolved rapid rotators with low nitrogen enrichment (Group~1) may be produced by close binaries
with an initial period small enough to ensure tidal locking and slow rotation but large
enough for highly non-conservative mass transfer, 
i.e. with an amount of mass accreted 
which is sufficient to spin-up the star but insufficient to enrich it significantly
\citep{pet05}. However, for many of the evolved stars in Group~1,
there is no indication of binarity. 
The single star nature for these objects would pose a serious challenge to the 
theory of rotational mixing.

While binaries 
might be able to populate Group~2 \citep{lan08}, another explanation may be more likely.
\citet{mor06} have analyzed ten slowly rotating Galactic early $\beta$-Cephei B-type stars
and found a highly enriched group (four out of ten) of which three have detected
magnetic fields. Although pulsations in these stars may
cause an apparent surface enrichment, \citet{bou07} report that only slight nitrogen enrichments
would be expected. As such, assuming that the effect of pulsations is negligible, 
there appears to be a correlation between magnetic fields and nitrogen enrichment.
Indeed, \citet{wol07} have attributed the large number of slow 
rotators often seen in massive star populations to 
magnetic locking of the star to the accretion disk during the star formation process and \citet{ale07}
also show a correlation between slow rotation and magnetic fields for magnetic A- and B-type stars.
Hence we postulate that the highly enriched slow rotators in the LMC are analogs
of the enriched Galactic magnetic stars. This would imply that, independent of metallicity, 
a significant fraction of early B-type stars are intrinsically magnetic, analogous to 
the well known situation for lower mass stars. In this context, the 
identification of three He-rich slowly rotating OVz stars in NGC\,346 \citep{mok06} suggests
that the phenomenon of intrinsic magnetic fields in massive stars may not be confined to the
B-star regime. If these magnetic fields are of fossil origin, 
i.e. possessing long-term stability \citep{bra04}, in contrast
to the fields produced by the \citep{spr02} mechanism,
one may speculate that stars in Group~2, and their Galactic counterparts, 
might be the progenitors of magnetars, analogous to the suggestion of \citet{fer06} 
that magnetic white dwarfs evolve from  magnetic A, F and late B stars.
While the fossil field hypothesis might well explain the slow rotation, 
the physical process which leads to the enrichment of nitrogen in
the stars of Group~2 remains to be identified.

The nitrogen abundances of the supergiants (Fig.~\ref{f_nvsini}(b))
appear to fall into two distinct groups, one group having a level of nitrogen enhancement
consistent 
with that seen for the majority of the core hydrogen burning objects ($<$7.2\,dex; Group 3), 
and a second having a much greater level of enrichment ($>$7.6\,dex; Group 4). It should be
noted that although the Group~4 objects appear to be
well fitted by evolutionary tracks with initial rotational
velocities of 200-300\,km\,s$^{-1}$, 
such a rotational velocity distribution for supergiants is inconsistent
with that for core-hydrogen burning objects and rotation does not increase the blue-supergiant
lifetime. Hence, the enrichments of the supergiants in Group~4  (70\% of the
non-binary supergiant sample) are incompatible with the
theory of rotational mixing. Additionally it should be noted that the lowest gravity stars
($<$2.8\,dex) are all enriched in nitrogen, implying that some evolutionary process not
accounted for in the models is responsible for the enrichment.

The simplest way to interprete this would be 
to characterize Group~3 as pre-red supergiant objects, and
Group~4 as post-red supergiant objects. 
The nitrogen abundances and rotational velocities in Group~4 are indeed consistent with the 
predictions for a blue loop stage \citep{heg00}. However, we note that our
models do not return to the high effective temperatures 
of the stars in Group~4 and hence their position on the H-R diagram
cannot be reproduced. As several of the enriched objects show evidence of binarity, 
mass transfer may also be important \citep{wel01}. 

To conclude, our study can not provide unambigous evidence for rotational mixing
acting in massive stars. The fast rotators in our sample can be interpreted in two ways.
Firstly as rotationally mixed stars, if the relatively non-enriched 
stars (Group~1) can be explained by
binary effects. However, if it were confirmed that these Group~1 stars 
are single stars, then the enriched rapid rotators may need to be 
understood as binary products --- and rotational mixing would be inefficient,
or much more complex than described by the present-day shellular models.
Additionally, the population of intrinsically slow rotators with nitrogen enhancements 
of up to a factor of $\sim$6 implies that studies of nitrogen enhancements with stellar samples
which are restricted to low projected rotational velocities --- which applies to most
previous works --- are not suited for constraining, or motivating, the adoption of 
rotational mixing in massive stars. Finally our supergiant sample implies that
these stars cannot be considered as representative of the amount of mixing that occurs
during the hydrogen burning main-sequence and hence should not be used to constrain this
process. 
In summary, our study provides a challenge to rotational mixing, and provides evidence
for two other processes affecting the rotation and the surface abundances of 
populations of massive stars, likely binarity and magnetic fields.


\acknowledgments

We are grateful to staff from the European Southern Observatory for assistance in
obtaining the data, and to STFC (UK Science and Technology Facilities
Council) and NWO (Netherlands Science Foundation)
for financial support. IH acknowledges financial support from 
DEL (Dept. of Employment \& Learning
Northern Ireland). SJS acknowledges the European Heads of Research Councils and European 
Science Foundation EURYI Awards scheme and the EC Sixth Framework Programme. We would
also like to thank the referee for highly constructive comments which significantly
improved this manuscript.

{\it Facilities:} \facility{ESO (VLT-FLAMES, 2.2m-WFI)}.

\newpage

\begin{thebibliography}{}
\bibitem[Abt et al.(2002)]{abt02} Abt, H.A., Levato, H. \& Grosso, M. 2002,
\apj, 573, 359
\bibitem[Alecian et al.(2007)]{ale07} Alecian, E., Wade, G.A., Catala, C. et al. 2007, 
To appear in the proceedings of `SF2A-2007: Semaine de l'Astrophysique Francaise', ed. J. Bouvier,
A. Chalabaev, \& C. Charbonnel, astro-ph/0710.1780 
\bibitem[Asplund et al.(2005)]{asp05} Asplund, M., Grevesse, N. 
\& Sauval, A.J. 2005, in
`Cosmic Abundances as Records of Stellar Evolution and Nucleosynthesis',
ed. T.G. Barnes III \& F.N. Bash, ASP Conf. Ser., 336, 25
\bibitem[Bourge et al.(2007)]{bou07} Bourge, P.-O., Th\'{e}ado, S. \& Thoul, A. 2007, 
Communications in Astroseismology, 150, 203
\bibitem[Braithwaite \& Spruit(2004)]{bra04} Braithwaite, J. \& Spruit, H.C. 2004, \nat, 431, 819
\bibitem[Brott et al.(in prep)]{bro07} Brott, I. et al. 2007, in prep.
\bibitem[Daflon et al.(2001)]{daf01} Daflon, S., Cunha, K., Butler, K. \& Smith, V.V. 2001, ApJ, 563 325
\bibitem[Dufton et al.(2005)]{duf05} Dufton, P.L., Ryans, R.S.I., 
Trundle, C., et al. 2005, \aap, 434, 1125
\bibitem[Evans et al. (2006)]{eva06} Evans, C.J., Lennon, D.J., 
Smartt, S.J. \& Trundle, C. 2006, \aap, 456, 623
\bibitem[Evans et al.(2005, 2006)]{eva05} Evans, C.J., Smartt, S.J.,
Lee, J.-K., et al. 2005, \aap, 437, 467
\bibitem[Ferrario \& Wickramasinghe(2006)]{fer06} Ferrario, L. \& Wickramsinghe, D. 2006, MNRAS, 367, 1323
\bibitem[Huang \& Gies(2006a)]{hua06} Huang, W. \& Gies, D.R. 2006a, \apj, 648, 580
\bibitem[Huang \& Gies(2006b)]{hua06b} Huang, W. \& Gies, D.R. 2006b, \apj, 648, 591
\bibitem[Heger \& Langer(2000)]{heg00} Heger, A. \& Langer, N. 2000, \apj, 544, 1016
\bibitem[Heger et al.(2000)]{heg00b} Heger, A., Langer, N. \& Woosley, 2000,\apj, 528, 368 
\bibitem[Hubeny \& Lanz(1995)]{hub95} Hubeny, I. \& Lanz, T. 1995, \apj, 439, 875
\bibitem[Hunter et al.(2007a)]{hun07a} Hunter, I., Dufton, P.L., 
Smartt S.J., et al. 2007a, \aap, 466, 277, Paper\,I
\bibitem[Hunter et al.(2007b)]{hun07b} Hunter, I., Lennon, D.J., 
Dufton, P.L., et al. 2007b, \aap, submitted, Paper III
\bibitem[Korn et al.(2002)]{kor02} Korn, A.J., Keller, S.C., Kaufer, A., et al.
2002, A\&A, 385, 143 
\bibitem[Langer (2008)]{lan08} Langer, N. 2008, in: IAU~Symp.~250, in press
\bibitem[Maeder \& Meynet(2001)]{mae01} Maeder, A. \& Meynet, G.
2001, \aap, 373, 555
\bibitem[Meynet \& Maeder(2005)]{mey05} Meynet, G. \& Maeder, A. 2005, \aap, 429, 581
\bibitem[Mokiem et al.(2006)]{mok06} Mokiem, M.R., de Koter, A., Evans, C.J., et al. 2006, \aap, 456, 1131
\bibitem[Morel et al.(2006)]{mor06} Morel, T. Butler, K., Aerts, C., 
  Neiner, C. \& Briquet, M. 2006, \aap, 457, 651
\bibitem[Peimbert et al.(2007)]{pei07} Peimbert, M., Luridiana, V. \& Peimbert, A. 2007, \apj, astro-ph/0701580
\bibitem[Petrovic et al.(2005)]{pet05}Petrovic, J., Langer, N., van der Hucht, K.A. 2005, \aap, 435, 1013
\bibitem[Spruit(2002)]{spr02} Spruit, H.C. 2002, \aap, 381, 923
\bibitem[Spruit(2006)]{spr06} Spruit, H.C. 2006, astro-ph/0607164
\bibitem[Trundle et al.(2007)]{tru07} Trundle, C., Dufton, P.L., 
Hunter, I., et al. 2007, \aap, 471, 625
\bibitem[Venn(1999)]{ven99} Venn, K.A. 1999, ApJ, 518, 405
\bibitem[Vink \& de Koter(2005)]{vin05} Vink, J.S. \& de Koter 2005, A\&A, 442, 587
\bibitem[Vink et al.(2001)]{vin01} Vink, J.S., de Koter, A. \& Lamers, H.J.G.L.M 2001, A\&A,
  369, 574
\bibitem[Wellstein et al.(2001)]{wel01} Wellstein, S., Langer, N.
\& Braun, H. 2001, \aap, 369, 939
\bibitem[Wolff et al.(2007)]{wol07} 
  Wolff, S.C, Strom, S.E., Dror, D. \& Venn, K. 2007, AJ, 133, 1092
\bibitem[Yoon \& Langer(2005)]{yoo05} Yoon, S.-C. \& Langer, N. 2005, \aap, 443, 643
\bibitem[Yoon et al.(2006)]{yoo06} Yoon, S.-C., Langer, N. \& Norman, C. 2006, \aap, 460, 199
\end{thebibliography}
\end{document}